\documentclass[useAMS, usenatbib]{mn2e}

\newcommand{\aap}{{\it A\&A}}

\newcommand{\nat}{{\it Nature}}
\newcommand{\apj}{{\it ApJ}}
\newcommand{\apjl}{{\it ApJL}}

\newcommand{\mnras}{{ \it MNRAS}}

\newcommand{\icarus}{{\it Icarus}}

\usepackage{parskip}
\usepackage{graphicx}
\usepackage{amssymb}
\usepackage{bbold}
\usepackage{hyperref}
\usepackage{empheq}
\usepackage{blindtext}
\usepackage{epsfig}
\usepackage{amsmath}
\usepackage{color}
\usepackage{comment}
\usepackage{txfonts}
\usepackage[update,prepend]{epstopdf}
\usepackage{float}
\usepackage{deluxetable}

\title{Outer-planet scattering can gently tilt an inner planetary system}

\author[Gratia and Fabrycky]{
Pierre Gratia,$^{1}$\thanks{E-mail: pierregratia@gmail.com (PG)} Daniel Fabrycky$^{2}$\thanks{E-mail: fabrycky@uchicago.edu (DF)}\\
$^{1}$Department of Physics, University of Chicago, Chicago, IL 60637, USA\\
$^{2}$Department of Astronomy and Astrophysics, University of Chicago, 5640 S. Ellis Ave., Chicago, IL 60637, USA
}

\date{Accepted XXX. Received YYY; in original form ZZZ}

\pubyear{2016}

\begin{document}
\label{firstpage}
\pagerange{\pageref{firstpage}--\pageref{lastpage}}

\maketitle
\begin{abstract}
Chaotic dynamics are expected during and after planet formation, and a leading mechanism to explain large eccentricities of gas giant exoplanets is planet-planet gravitational scattering.  The same scattering has been invoked to explain misalignments of planetary orbital planes with respect to their host star's spin.  However, an observational puzzle is presented by Kepler-56, which has two inner planets (b and c) that are nearly coplanar with each other, yet are more than 45 degrees inclined to their star's equator. Thus the spin-orbit misalignment might be primordial. Instead, we further develop the hypothesis in the discovery paper, that planets on wider orbits generated misalignment through scattering, and as a result gently torqued the inner planets away from the equator plane of the star.  We integrated the equations of motion for Kepler-56 b and c along with an unstable outer system initialized with either two or three Jupiter-mass planets. We address here whether the violent scattering that generates large mutual inclinations can leave the inner system intact, tilting it \emph{gently}. In almost all of the cases initially with two outer planets, either the inner planets remain nearly coplanar with each other in the star's equator plane, or they are scattered violently to high mutual inclination and high spin-orbit misalignment. On the contrary, of the systems with three unstable outer planets, a spin-orbit misalignment large enough to explain the observations is generated 28\% of the time for coplanar inner planets, which is consistent with the observed frequency of this phenomenon reported so far. We conclude that multiple-planet scattering in the outer parts of the system may account for this new population of coplanar planets hosted by oblique stars. \vspace{0.3 in}
\end{abstract}


\section{Introduction}

As part of the great diversity of known planetary systems, hot Jupiters are frequently observed with spin-orbit misalignment \citep{2008Hebrard, fabrycky09, triaud10, morton11, moutou11, albrecht12, hebrard13}. Planets with even slightly more widely spaced orbits have only rarely allowed spin-orbit measurement, due to observational difficulties. 

Kepler-56 belongs to the few discovered systems that contain several, more widely-spaced planets (planets b and c have periods 10.5 days and 21.4 days), and also a spin-orbit measurement.  In fact, it was the first such system to show spin-orbit misalignment \citep{2013Huber}. Both inner planets are misaligned with their host star's spin axis by at least $45^\circ$, while being mutually aligned to within about $10^\circ$.  The geometry of this system, as well as its eventual fate, has been detailed by \cite{2014Li}. 

In Kepler-56, \cite{2013Huber} also found a radial acceleration consistent with a third giant planet (call it planet d) in a several-AU orbit. That detection inspired them to propose the following scenario to explain the misalignment (following \citealt{2010Mardling} and \citealt{2011Kaib}).  Suppose that a fourth planet is initially in the outer parts of the system, and all planets and the stellar spin are coplanar to within a few degrees.  The orbits of the two outer planets may go unstable, initiating an epoch of gravitational scattering.  Eventually planet d could eject the additional planet, leaving planet d on an eccentric orbit, with an inclination $i_d$.  Thus the chaotic dynamics could begin with a relatively flat system of planets and inject inclination into its outer parts.  Two-planet scattering simulations leave behind a planet with an inclination at or above $i_0 = 22.5^\circ$ about 1\% of the time \citep{2001Ford}, whereas three-planet simulations do so $\sim 30$\% of the time \citep{2008Chatterjee}.  From then on, this inclined outer planet d would slowly cause the inner planets to precess, periodically sampling spin-orbit angles between $0$ and $2 i_d$, assuming planet d's angular momentum dominates the rest of the system.  That tilt would be ``gentle'' in the sense that the inner planets would maintin coplanarity \citep{1997Innanen, 2011Kaib, 2013Huber, Boue2014}.  In the specific case of Kepler-56, the inner planets also have low eccentricities ($<0.1$) as determined by the transit timing variation analysis \citep{2013Huber}, further evidence that they did not directly participate in the scattering.

Interestingly, similar dynamics have recently been noted in the Solar System, supposing the solar obliquity is due to a distant perturbing planet \citep{2016Gomes,2016Bailey}, a revisitation of an old idea \citep{1972GoldreichWard}.

Given that spin-orbit misalignment seems to be a generic feature of different kinds of planetary systems, it is of great importance to understand the mechanism(s) that can lead to such an outcome. The weakest part of the scattering scenario seems to be the need for the scattering planets to leave the inner planets undisturbed.  This aspect can only be checked via numerical simulations. 

The plan for this paper is as follows.  In section~\ref{sec:method}, we describe the suite of numerical simulations: the method and initial conditions. In section~\ref{sec:ex}, we give a few examples that lead to ejections or collisions of the outer planets, yielding a system with misaligned inner planets. Section~\ref{sec:res} will be dedicated to the statistical outcomes, and their interpretation. We will discuss absolute inclinations as well as mutual inclinations. Finally, we conclude the paper with a summary of our results in section~\ref{sec:concl}.

\section{Scattering simulations for the Kepler-56 system} \label{sec:method}

We investigate the scattering hypothesis for the case of four or five initial planets, and see if it can lead to spin-orbit misalignment for a system such as Kepler-56. We simulate the dynamics of the system for three sets of initial parameters, as discussed below.  
Events such as an ejection or collision of outer planets may leave us with a distant planet to create the observed radial velocity trend \citep{2013Huber} and subsequently produce spin-orbit misalignment of the inner two planets. However, simulated systems that retain their outer planets on calm orbits will not create any significant misalignment; we do not record them in the plots of section $5$. 

For our simulations, we use \textit{Mercury} \citep{1999C}, an N-body simulation code. The Burlisch-Stoer integrator is used with an accuracy parameter of $10^{-12}$. Collisions between planets are assumed to result in a perfect merger. Beyond a critical distance, a planet is considered ejected, and is taken out of the integration.  For the two outer-planet runs, 100 AU is used; for the three outer-planet runs, 1000 AU is used.  In all integrations, the host star mass $M_{star} = 1.32 M_{sun}$.

For planetary masses and radii, for models with the inner planets, we took values corresponding to Kepler-56b and Kepler-56c (see Table 1 in \citealt{2013Huber}).  For the outer planets, we took one Jupiter mass and 3 Jupiter radii, representing their radius at an early age \citep{2001Burrows}.

We assume that on scattering timescales, the star's spin orientation will not change. Even on secular timescales, this assumption seems justified (e.g., \citealt{2013Huber,Boue2014}).  In this work, we assume that the stellar spin remains oriented perpendicular to the initial plane of the planets, from which their inclinations are measured.  Thus we interpret final inclinations to be equivalent to spin-orbit misalignment angles, and we forgo modelling the stellar spin.

The initial inclinations were chosen to be: 
\begin{itemize}
\item Zero for the outermost planet;
\item A uniform-random number chosen from the interval $[0,5^\circ]$ for the others. 
\end{itemize}
Thus we choose an almost planar configuration for all planets. The initial angular parameters --- argument of periastron $\omega$, nodal angle $\Omega$, and mean anomaly $M$ --- were chosen randomly between $[0,2\pi]$ for the outer planets, but fixed for the inner two. Kepler-56b and c's initial angular parameters are from table S6 of \cite{2013Huber}, which we repeat for convenience -- as well as summarizing the other intial conditions -- in Table~\ref{tab:init} and Table~\ref{tab:init5}.

The mutual inclination $I$, a crucial parameter to test the scattering hypothesis for two misaligned planets, is defined with the orbital node $\Omega$:

\begin{equation}
\cos I = \cos i_b  \cos i_c + \sin i_b  \sin i_c  \cos\Omega.
\end{equation}

The initial eccentricities were chosen randomly within $[0,0.01]$ for all planets. 

We ran three sets of simulations. 

The first set of simulations were 173 runs with two inner planets with Kepler-56 b and c properties, and two outer gas giants, called d and e.  Its properties are detailed in Table~\ref{tab:init}. 

We followed \cite{2001Ford} to initialize the outer planets. The initial parameters are:
\begin{itemize}
\item $\alpha\equiv a_d/a_e$ in the range $[0.769, 0.781]$, where $a_d = 5 AU$ ($\simeq$ Jupiter orbit) and $a_e$ are the semi-major axes of the two outer planets (see section $3$ of \cite{2001Ford} for a justification of this range);
\item Total integration time: $5.23\times 10^9$ days $ \simeq 1.43\times 10^7$ years $\simeq 1.47\times 10^6$ OUTER1 orbits;  \cite{2001Ford} integrate until $1.6\times 10^6$ Jupiter orbits, and we see from their figure 2 that the branching ratios are in place by $\simeq 10^6$ Jupiter orbits, at least for the two-planet case. 
\end{itemize}

The second set of simulations consists of 73 integrations with the same two outer planets as 73 of the first set, but excluding the two inner planets.  This is in order to check the rates of different outcomes. The reason for doing this was to (i) compare with the two-planet runs of \cite{2001Ford}, and (ii) see if the branching ratios of outer planet ejections, collisions, and stable systems, are different from the four-planet case. In principle, the inner planets could change the outcome probabilities by modifying the scattering (e.g., \citealt{2015Mustill}). 

\begin{table*}
\caption{Initial conditions for 4-planet simulations: set 1.}
\label{tab:init}
\centering 
\begin{tabular}{l | c | c | c | c | c | c | c | c }
\hline\hline 
Planet     & semi-major  & eccentricity & inclination & nodal angle & periapse angle & mean anomaly & mass & radius\\
name & axis $a$, [AU] & $e$ & $i$ [deg] & $\Omega$ [deg] & $\omega$ [deg] & $M$ [deg] & $M_{\rm Jupiter}$ & $R_{\rm Jupiter}$       \\
[0.5ex]
\hline 
Kepler-56 b & 0.1028 & [0.0,0.01] & [0,5] & 0.0 & 0.0 & 57.0 & 0.0695 & 0.581 \\
Kepler-56 c & 0.1652 & [0.0,0.01] & [0,5] & 0.0 & 0.0 & 182.0 & 0.569 & 0.875 \\
OUTER1 & 5.0000 & [0.0,0.01] & [0,5] & [0,360] & [0,360] & [0,360] & 1.0 & 3.0 \\
OUTER2 & [6.4020,6.5020] & [0.0,0.01] & 0.0 & [0,360] & [0,360] & [0,360] & 1.0 & 3.0 \\
\hline 
\end{tabular}
\end{table*}

A third set of simulations has three outer planets, as well as the inner planets representing Kepler-56 b and c.  From \cite{1996C}, we recognize that three planets can be spaced much wider than two planets and still go unstable.  We choose initial semi-major axes: $5.0000$~AU, $7.1220$~AU, and $10.1446$~AU, and the same random selection of other orbital elements as above, which resulted in a wide range of instability timescales.  We integrated $10^8$~yr for this set, and continued integrating even after the outer planet system became unstable.  See Table~\ref{tab:init5} for a summary of these initial conditions. 

\begin{table*}
\caption{Initial conditions for 5-planet simulations: set 3.}
\label{tab:init5}
\centering 
\begin{tabular}{l | c | c | c | c | c | c | c | c }
\hline\hline 
Planet     & semi-major  & eccentricity & inclination & nodal angle & periapse angle & mean anomaly & mass & radius\\
name & axis $a$, [AU] & $e$ & $i$ [deg] & $\Omega$ [deg] & $\omega$ [deg] & $M$ [deg] & $M_{\rm Jupiter}$ & $R_{\rm Jupiter}$       \\
[0.5ex]
\hline 
Kepler-56 b & 0.1028 & [0.0,0.01] & [0,5] & 0.0 & 0.0 & 57.0 & 0.0695 & 0.581 \\
Kepler-56 c & 0.1652 & [0.0,0.01] & [0,5] & 0.0 & 0.0 & 182.0 & 0.569 & 0.875 \\
OUTER1 & 5.0000 & [0.0,0.01] & [0,5] & [0,360] & [0,360] & [0,360] & 1.0 & 3.0 \\
OUTER2 & 7.1220 & [0.0,0.01] & [0,5] & [0,360] & [0,360] & [0,360] & 1.0 & 3.0 \\
OUTER3 & 10.1446 & [0.0,0.01] & 0.0 & [0,360] & [0,360] & [0,360] & 1.0 & 3.0 \\
\hline 
\end{tabular}
\end{table*}

\section{ Examples of the dynamical evolution } \label{sec:ex}
The examples that follow are from simulation sets 1 and 3, and they focus on cases where both inner planets have a final inclination larger than $10^\circ$ with respect to their host star's spin axis.  The final values are given in Table~\ref{tab:sys} and labelled by S1, S2, etc.

\subsection{Examples with three initial outer planets }
\label{sec:ex1}
\subsubsection{System S1 - secular excitation during scattering}
\begin{figure}  
\begin{center}  
\includegraphics[width=0.49\textwidth]{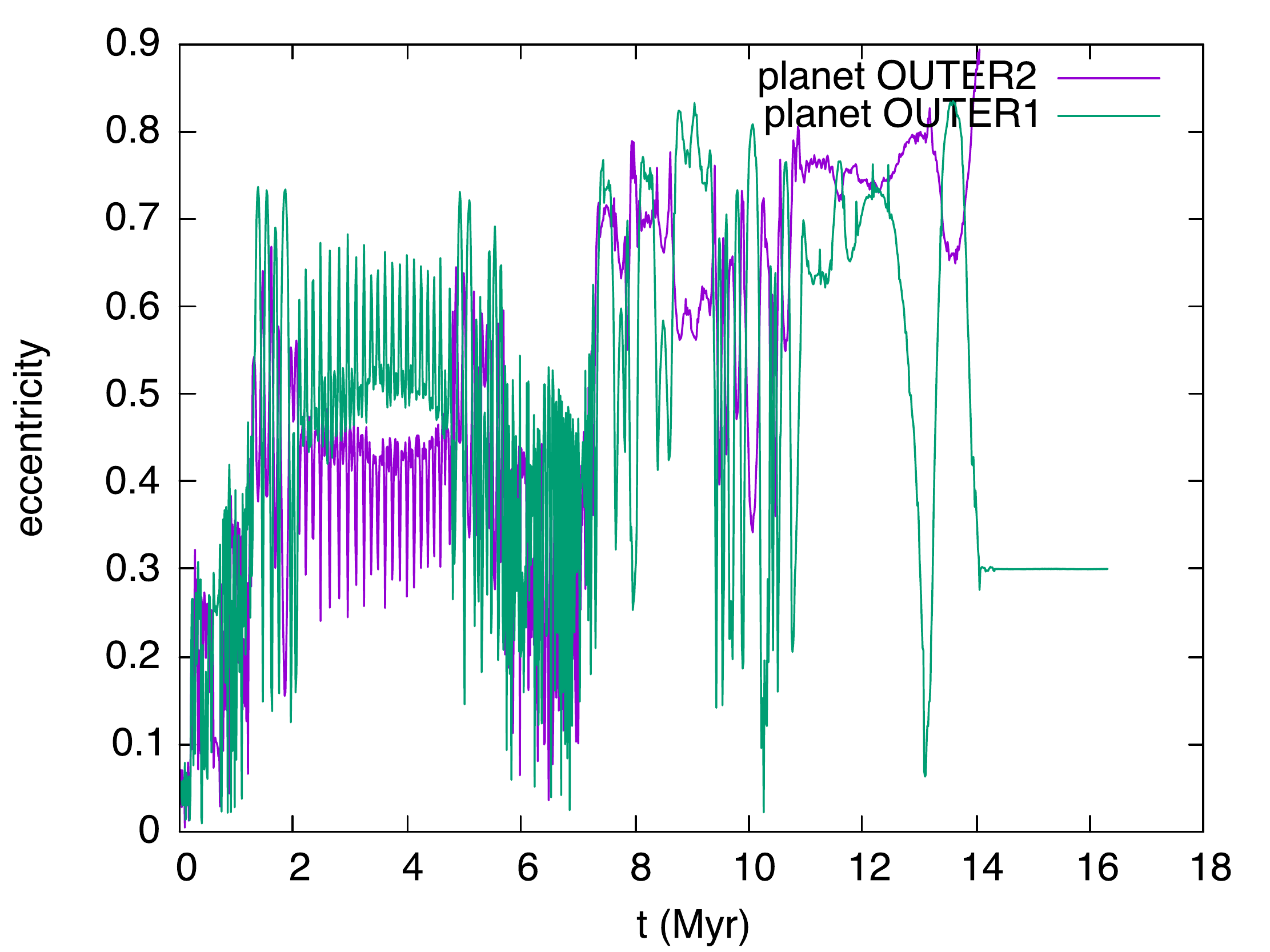} 
\caption{System S1. This plot shows the eccentricities over time of the outer two planets. Here, the ejection took place late in the simulation.}
\label{fig:ecc1}  
\end{center}  
\end{figure}

Here we have a common case where one of the outer two planets has been ejected, and we are left with a three planet system. Figure \ref{fig:ecc1} shows the eccentricities of the outer two planets over time. As we can see, the eccentricity of OUTER2 approaches unity and is ejected toward the end of the integration time.

\begin{figure} 
\begin{center}  
\includegraphics[width=0.49\textwidth]{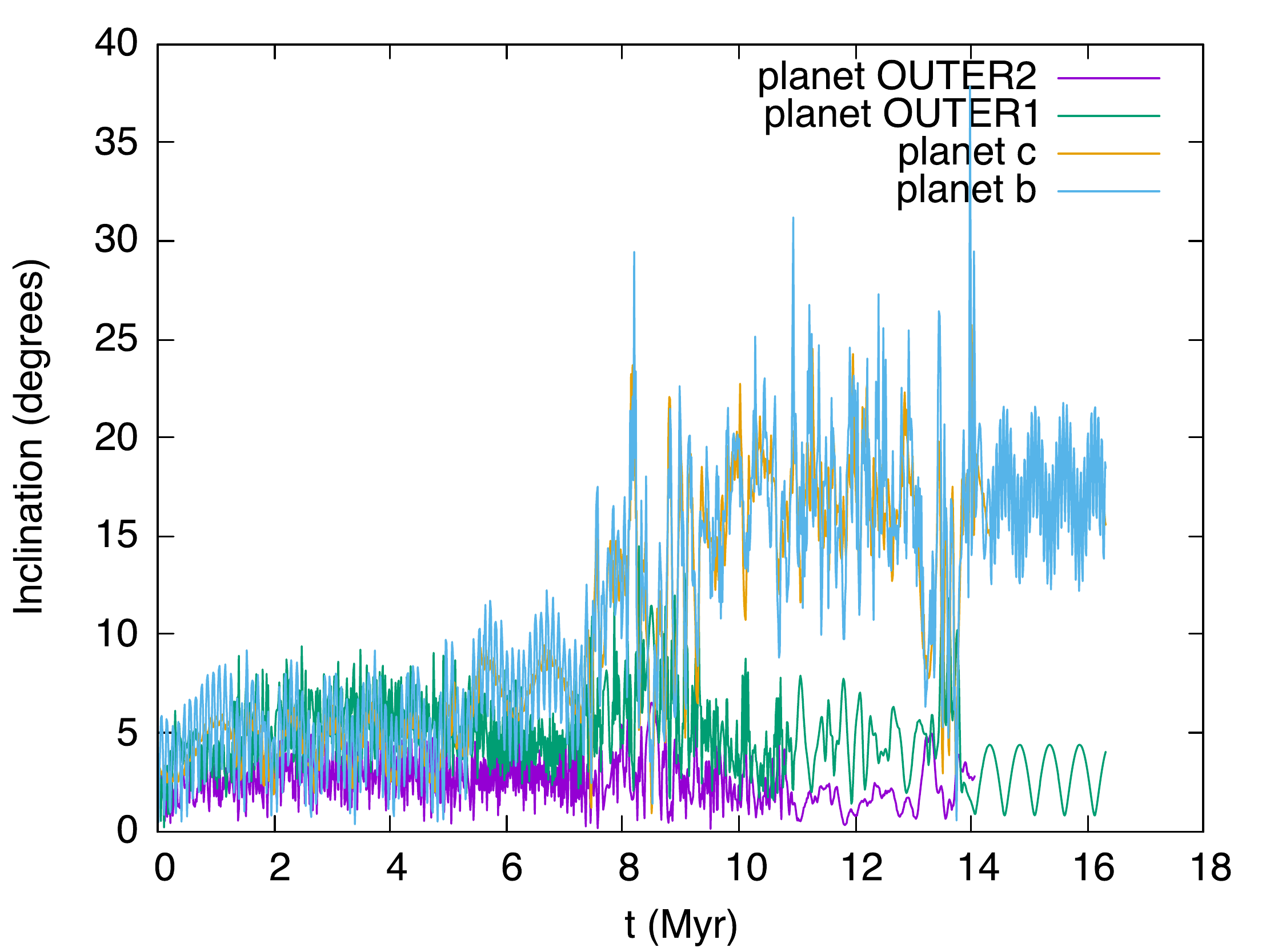} 
\caption{System S1. The absolute inclinations of the inner two planets over time vary strongly during the second half of the integration time. However, while one of them briefly reaches a $35^\circ$ inclination toward the end, the other never inclines by more than about $25^\circ$.}
\label{fig:inc1}  
\end{center}  
\end{figure}
\begin{figure}  
\begin{center}  
\includegraphics[width=0.49\textwidth]{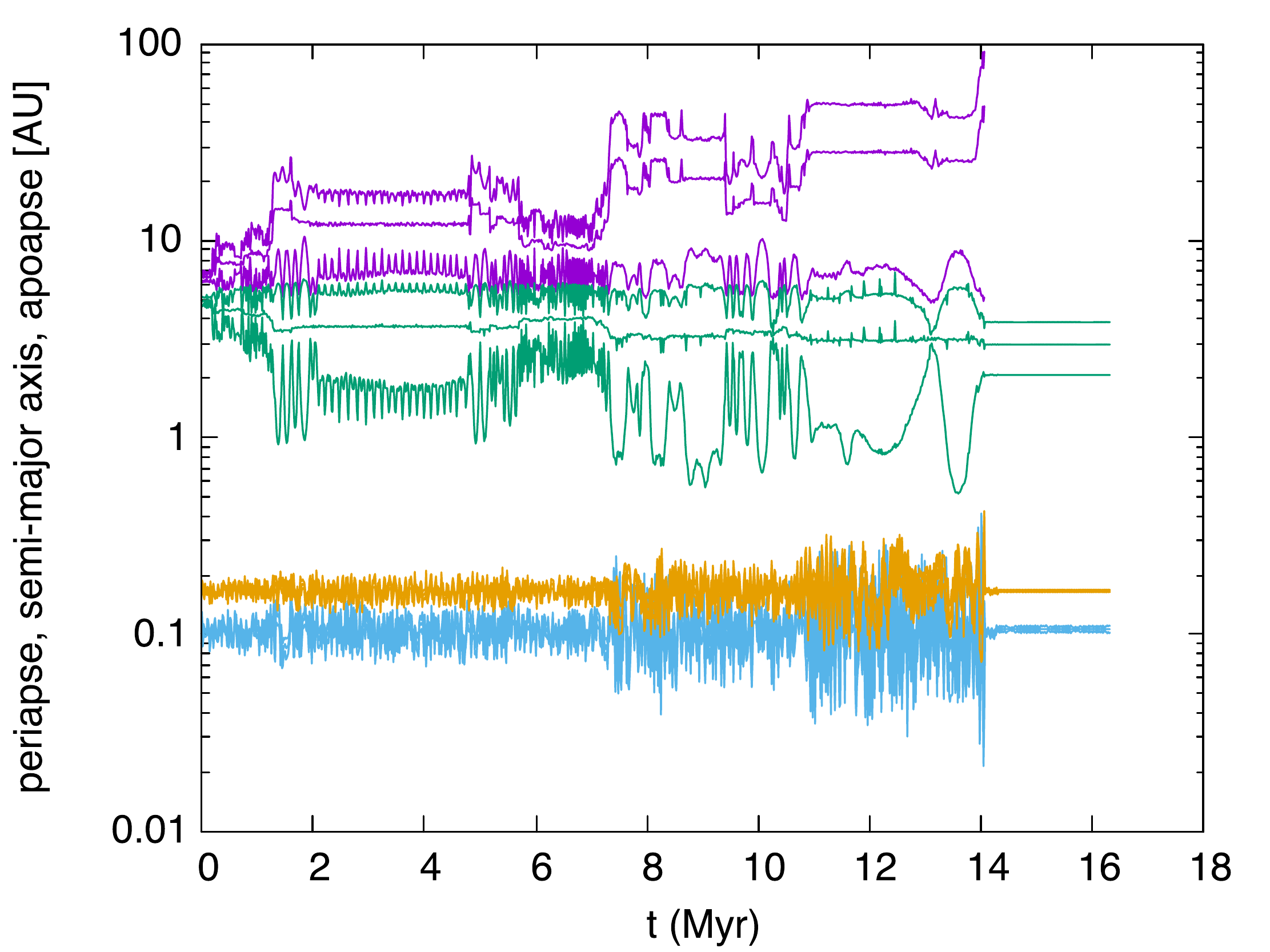} 
\caption{System S1. Aphelion, Perihelion, and semi-major axis development over time for a system that ejects its outermost planet.}
\label{fig:all1}  
\end{center}  
\end{figure}

 
Figure \ref{fig:inc1} shows the inclinations of the inner two planets. Interestingly, they fluctuate wildly in the second half of the integration time, but most of the time their values stay close to each other.  However the final inclinations are moderate, around $15^\circ$, as opposed to the greater than $45^\circ$ degrees that we observe for Kepler-56. We also plot the perihelion, aphelion, and semi-major axis for all four planets in figure~\ref{fig:all1}.

\subsubsection{System S2 - torquing following scattering}   
\begin{figure}  
\begin{center}  
\includegraphics[width=0.49\textwidth]{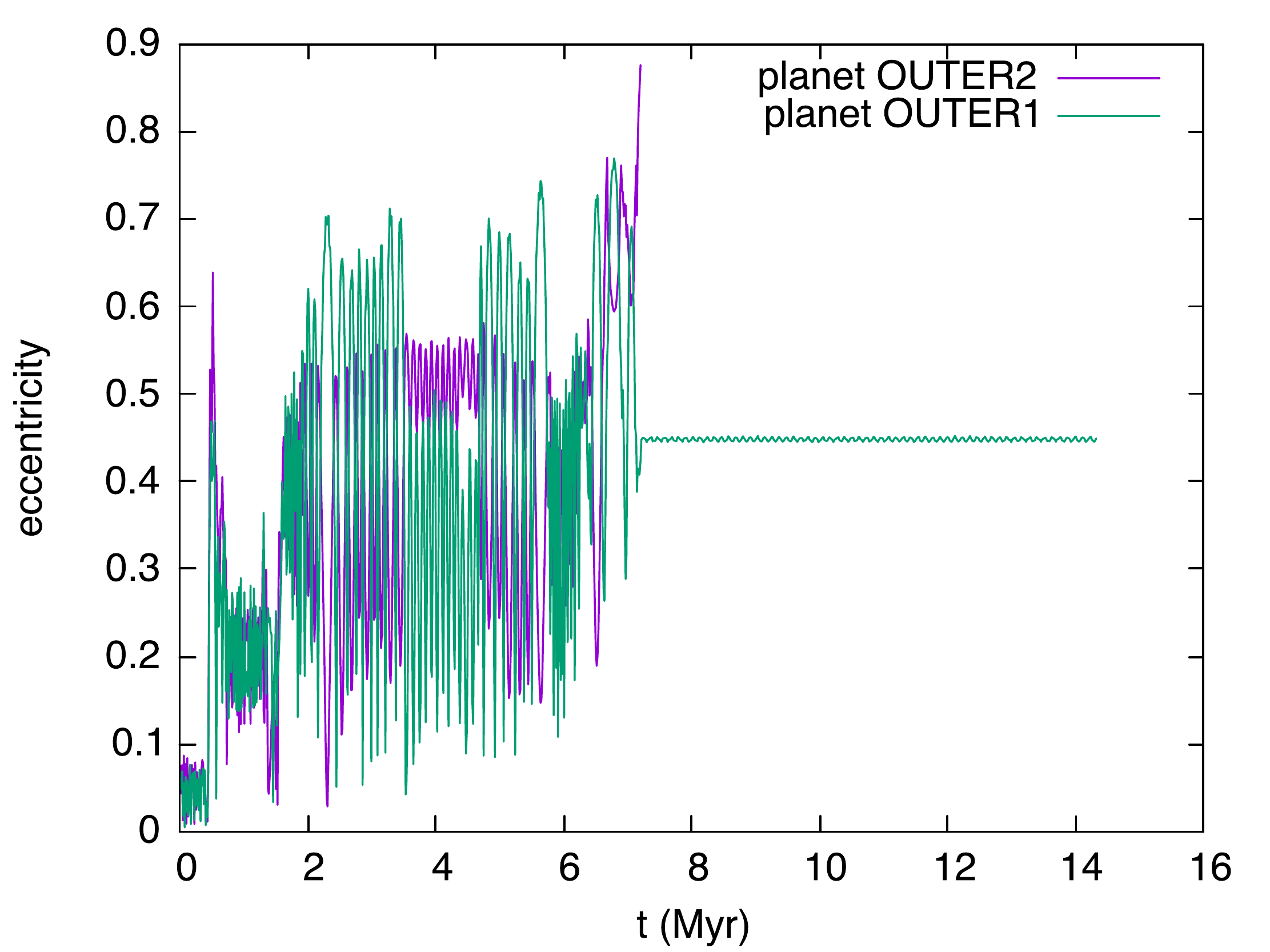} 
\caption{System S2. Another system's eccentricity development of its outer two planets. The outermost planet is ejected mid-way through the simulation.}
\label{fig:ecc2}  
\end{center}  
\end{figure}

\begin{figure}  
\begin{center}  
\includegraphics[width=0.49\textwidth]{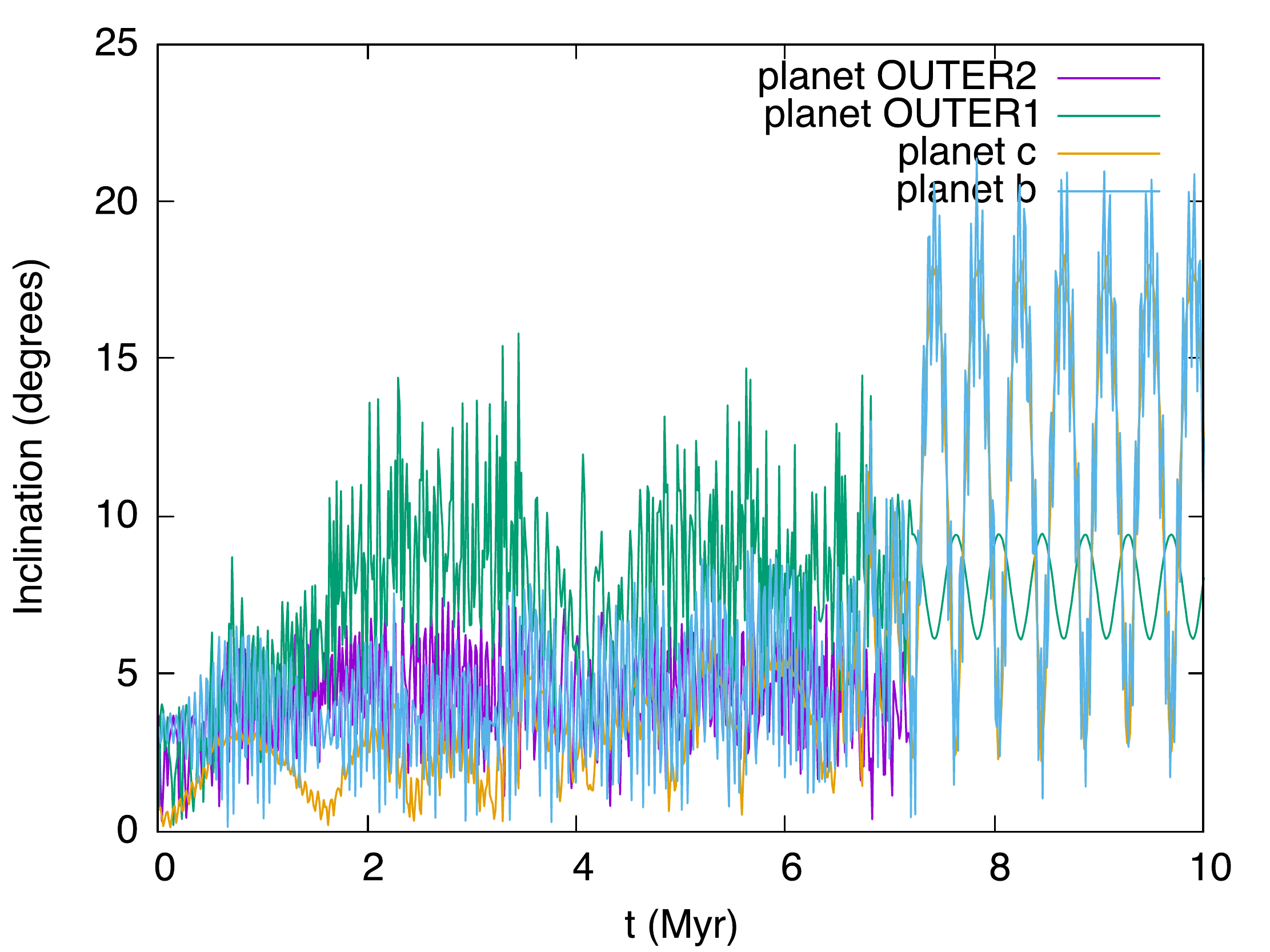} 
\caption{System S2. The absolute inclinations of inner two planets over time enter a periodic pattern whose maximal amplitude stays below $25^\circ$.}
\label{fig:inc2}  
\end{center}  
\end{figure}

Here we show a system where the ejection happened earlier, with the excitation of the outermost planet's eccentricity above 1 (fig.~\ref{fig:ecc2}). After the ejection occurred, the surviving outer planet has an inclination of $9^\circ$.  The inner planets' inclinations then enter a periodic pattern of inclinations between a few and $\sim20^\circ$ (fig.~\ref{fig:inc2}). Again, these values are well below the required $\sim 45^\circ$ for Kepler-56. On the other hand, the sequence of events, and the fact that the inner planets maintain a mutual inclination of $<5^\circ$, fulfills the spirit of the mechanism described by \cite{2013Huber}, even though it does not match the Kepler-56 system quantitatively.

\begin{figure}  
\begin{center}  
\includegraphics[width=0.49\textwidth]{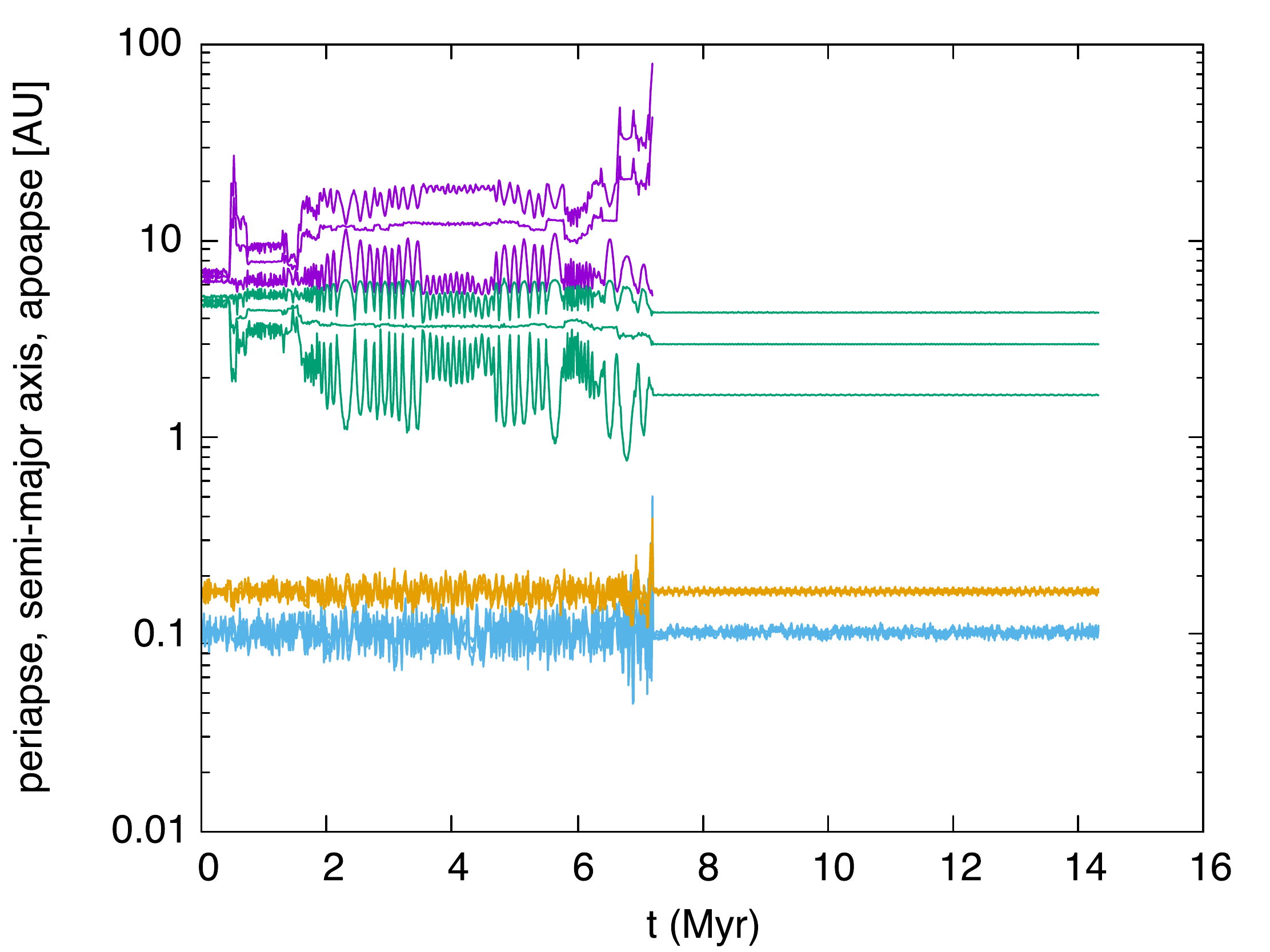} 
\caption{System S2. Aphelion, Perihelion, and semi-major axis development over time for a system that ejects its outermost planet.}
\label{fig:all2}  
\end{center}  
\end{figure}

The stability of the final configuration is evident in that the remaining three planet's perihelion and aphelion axes remain constant after the ejection (fig.~\ref{fig:all2}).

\subsubsection{System S3 - Ejection of second-outermost planet } \label{sec:s3}

\begin{figure} 
\begin{center}  
\includegraphics[width=0.49\textwidth]{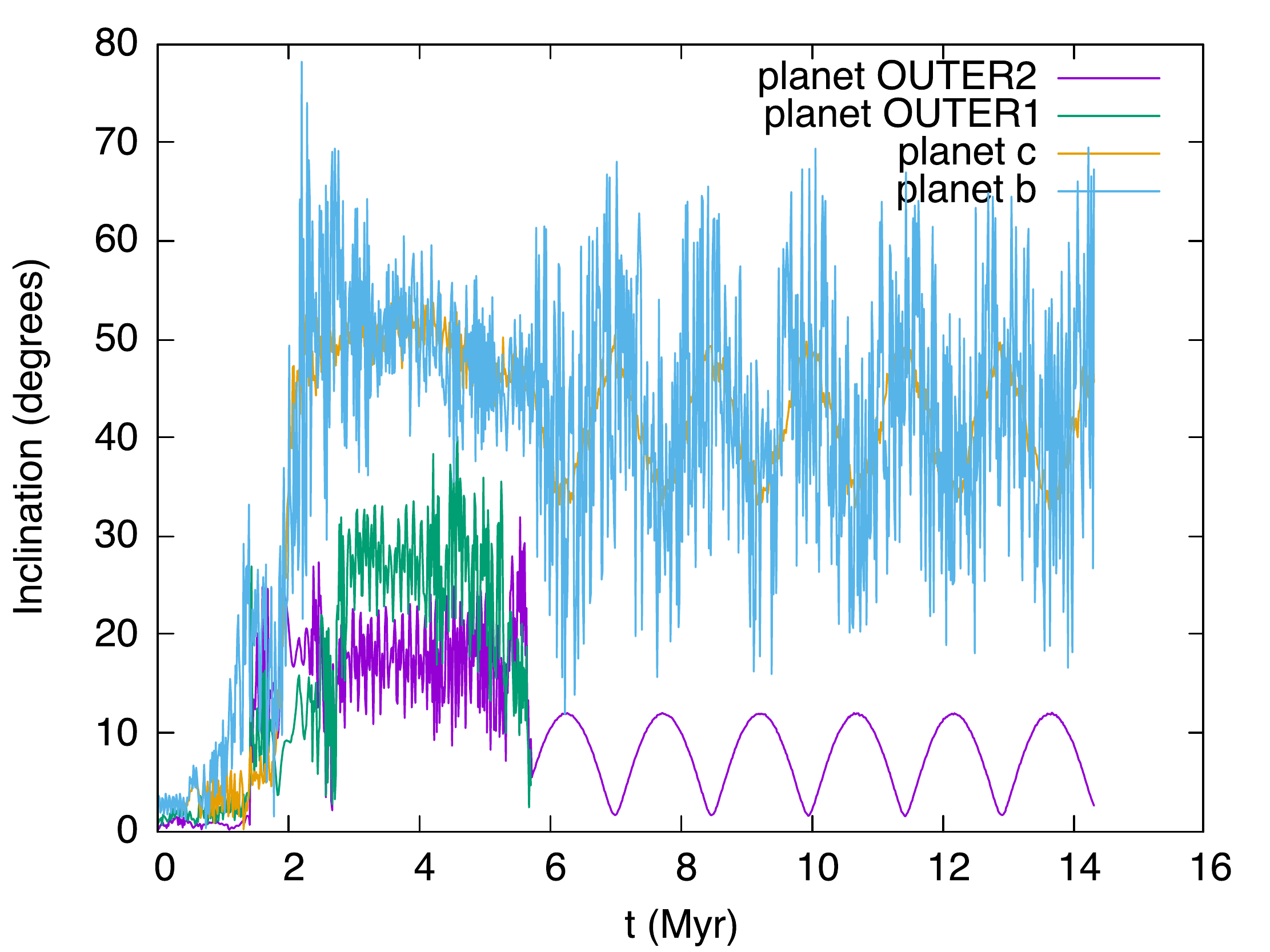} 
\caption{System S3. The absolute inclinations of the four planets over time. This is the only system we found that ended up with large inclinations of both inner planets, at $56^\circ$ and $47^\circ$, respectively (note the higher range on the y axis as compared to the previous two examples). However, the mutual inclination of $\sim 20^\circ$ is too large for considering this system a Kepler-56 candidate. }
\label{fig:inc3}  
\end{center}  
\end{figure}

\begin{figure}  
\begin{center}  
\includegraphics[width=0.49\textwidth]{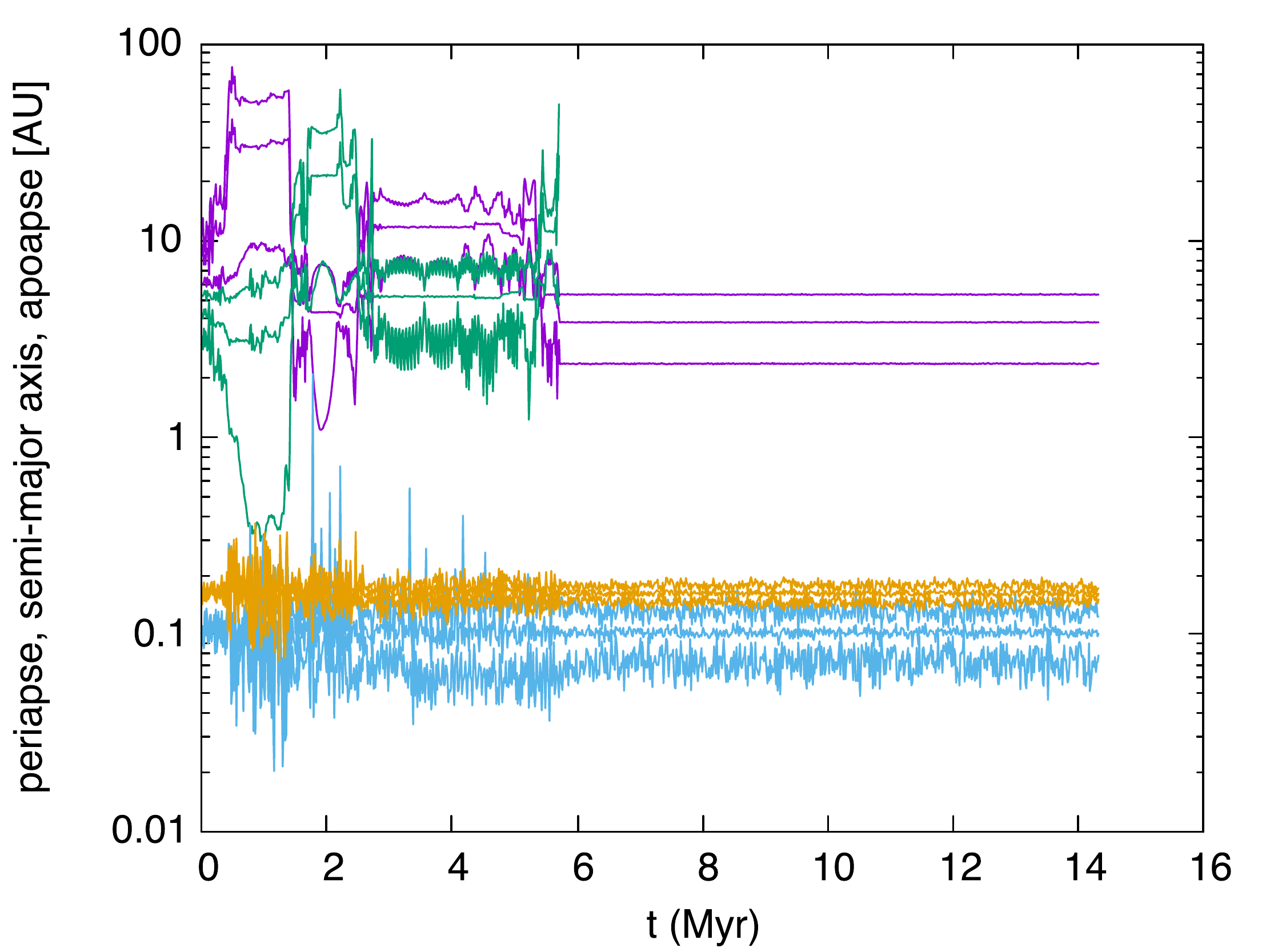} 
\caption{System S3. Development of semi-major, aphelion, and perihelion axes over time for the system that ejected its second-outermost planet.}
\label{fig:all3}  
\end{center}  
\end{figure}

We do not find many systems that ejected the second-outermost planet. Interestingly, it is such a system that produced the largest final inclinations for both inner planets we found in our simulations of two initial outer planets.  We attribute this to the complicated scattering history that allowed growth of large mutual inclination.  We omit plotting the outer planets' eccentricities; it looks very similar to Figure~\ref{fig:ecc2}. Figures \ref{fig:inc3} and \ref{fig:all3} show the inclinations and the orbital distances, respectively, of this system. While the second inner planet's inclination fluctuates between $30^\circ$ and $50^\circ$, the innermost planet's inclination reaches a minimum of about $10^\circ$ nd a maximum of over $70^\circ$. That greater variation is due to large mutual inclination (see section~\ref{sec:restwo}), which makes this simulated system a poor match to Kepler-56. 

\subsection{Examples with three initial outer planets} \label{sec:secoutej}
In the runs that start with three outer planets (simulation set 3, with five planets total), we see various outcomes: collisions between two outer-planets which either eject or retain the third outer-planet, and ejections of two outer-planets.  We did not see a merger of all three outer-planets.  Thus the outcomes of unstable systems were one or two outer-planets, with a survivor sometimes having twice the mass of the original planets. 

The different chaotic pathways leave a large variety of outcomes for the inner planets.  The inner planets excite to large inclination by various modes.  In some cases, during the scattering of the outer planets, the inner planets raise to a certain mutual inclination, and they subsequently do not evolve much in inclination.  In other cases, the scattering is fast compared to secular excitation, and it leaves the inclination of the inner two planets oscillating between nearly their original value and a maximum value of tens of degrees.  

\subsubsection{System S14 - Secular excitation during scattering}
\begin{figure}  
\begin{center}  
\includegraphics[width=0.49\textwidth]{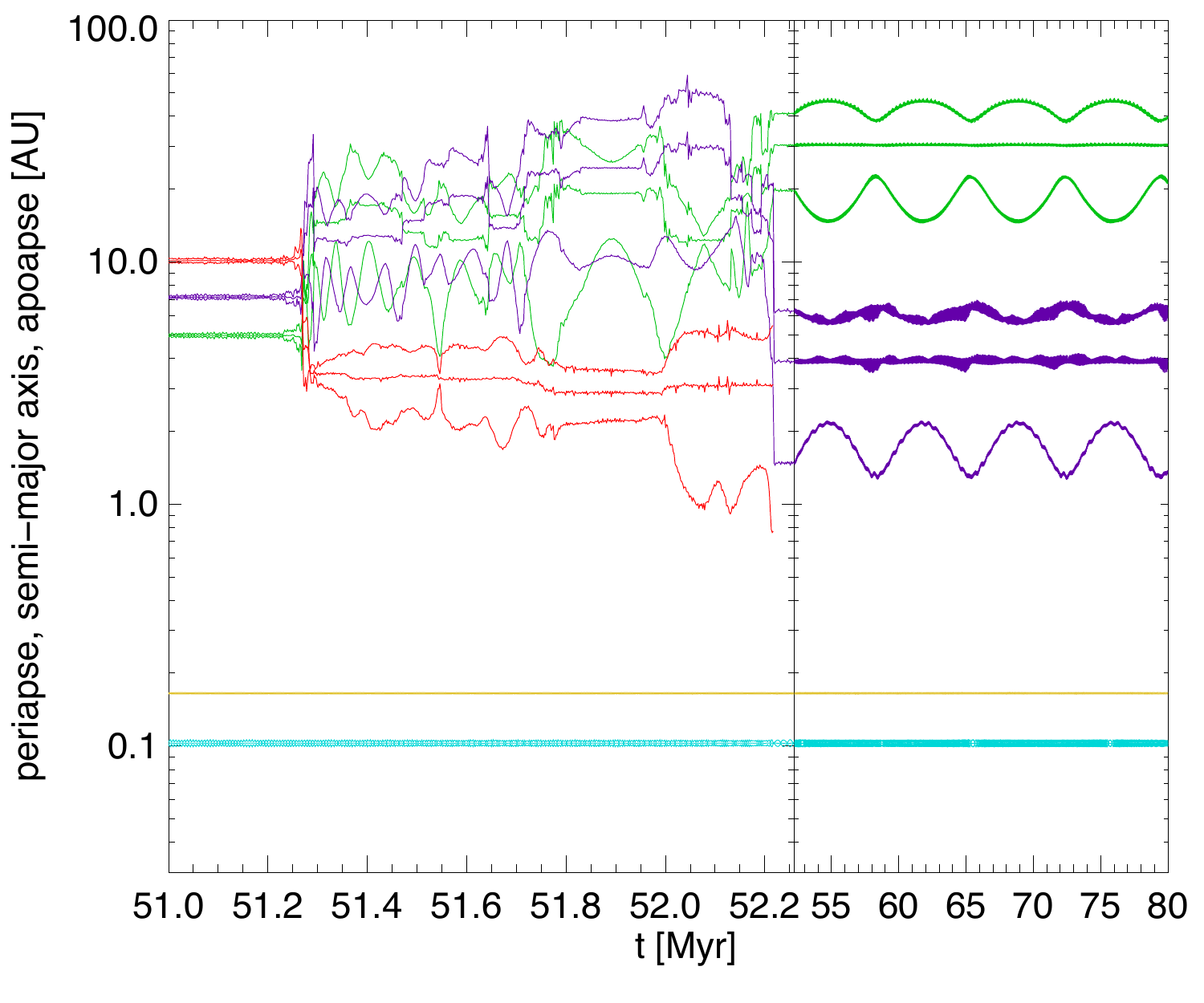} 
\caption{ System S14. Distances as a function of time: the semimajor axis, apastron distance, and periastron distance for each of the planets is shown, for a system in which the outer two planets merge after $\sim1$~Myr of scattering.  }
\label{fig:aqq28}  
\end{center}  
\end{figure}

\begin{figure}  
\begin{center}  
\includegraphics[width=0.49\textwidth]{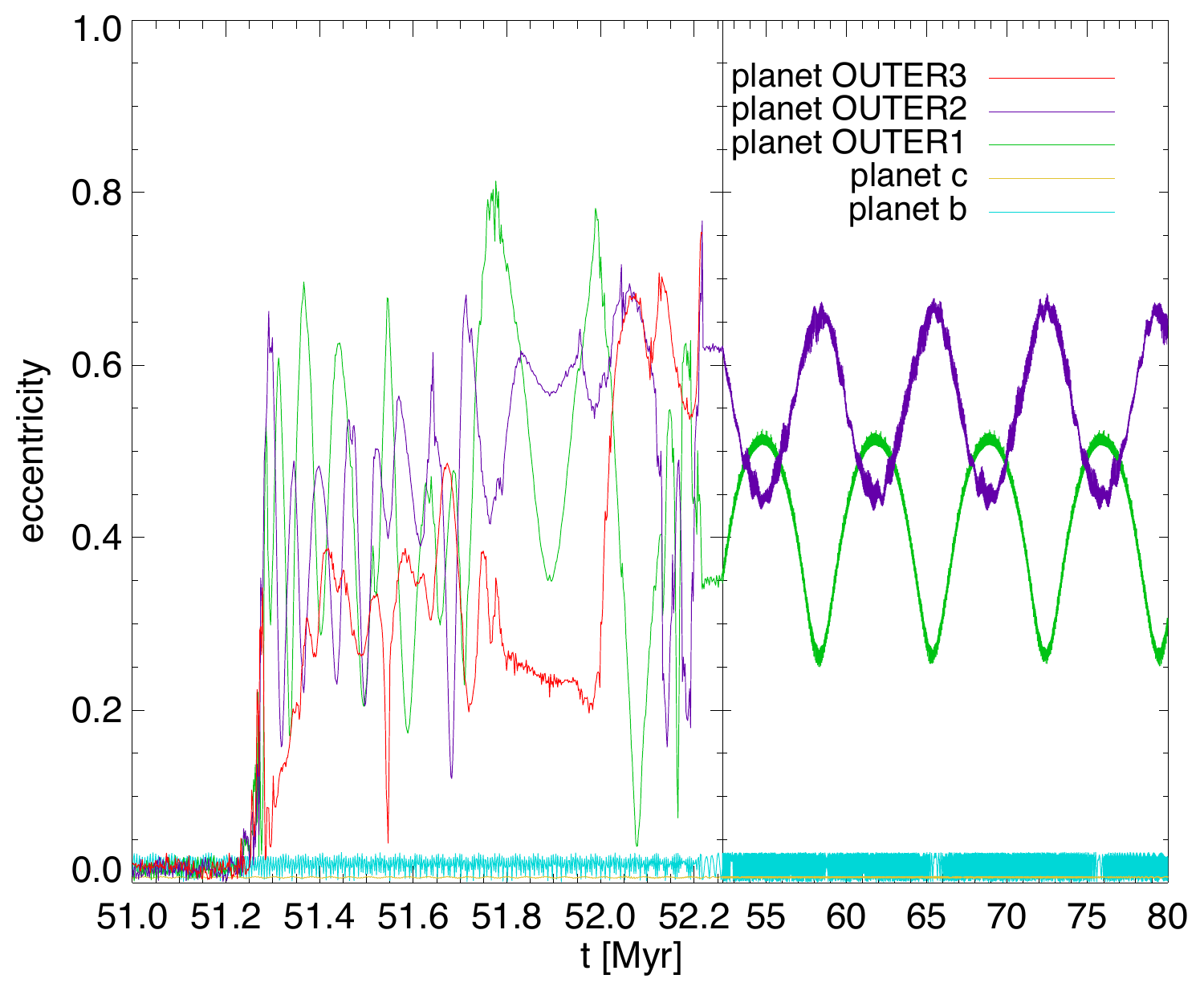} 
\caption{ System S14.  Eccentricity as a function of time.}
\label{fig:ecc28}  
\end{center}  
\end{figure}

\begin{figure}  
\begin{center}  
\includegraphics[width=0.49\textwidth]{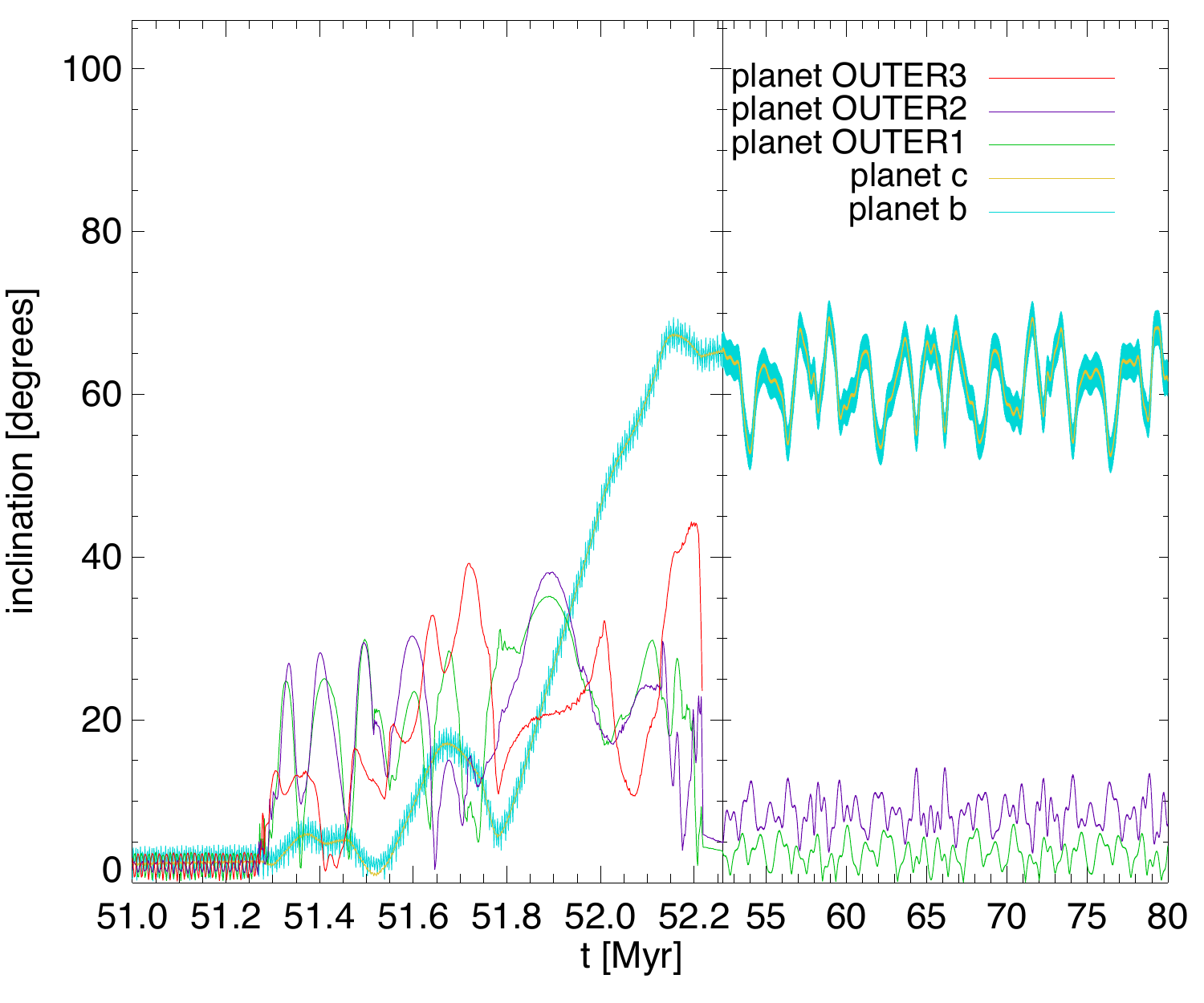} 
\caption{ System S14.  Inclination as a function of time. The inclination is relative to the original plane of the outermost planet.}
\label{fig:inc28}  
\end{center}  
\end{figure}

An example of secular excitation during scattering of the inner two planets is found in figures~\ref{fig:aqq28}, \ref{fig:ecc28}, and \ref{fig:inc28}. In this run, the scattering of three planets occurs over $\sim 1$~Myr.  For most of the scattering, the outermost planet of the three has been traded into the inner position of the three, whereas the other two planets have crossing, eccentric orbits.  The scattering ends when one of those two planets collides with the traded planet. The two resulting planets have separated enough orbits to no longer scatter, but they continue secular cycles of eccentricity.  The remnant outer planets have inclinations of $\lesssim 10^\circ$. 

During the scattering, the inner two planets (Kepler-56 b and c) never endure any scattering evolution (for instance, their eccentricities remain small; figure~\ref{fig:ecc28}) but are secularly excited to an inclination of $65^\circ$ (figure \ref{fig:inc28}).  Since the final inclinations of the outer planets are modest, the only indication of the past episode of scattering is the eccentricities of the outer two planets. 

The interesting concept here is that different parts of the planetary system react to perturbations differently; the outer planets can change their orbits on orbital timescales due to scattering, whereas the inner planets may only secularly change during this same evolution.  As far as we are aware, this is a new mechanism that has not been discussed before in spin-orbit alignment context; it is distinct from the hypothesis of \cite{2013Huber}. 


\subsubsection{System S15 - Impulsive Excitation and Ringing }
\begin{figure}  
\begin{center}  
\includegraphics[width=0.49\textwidth]{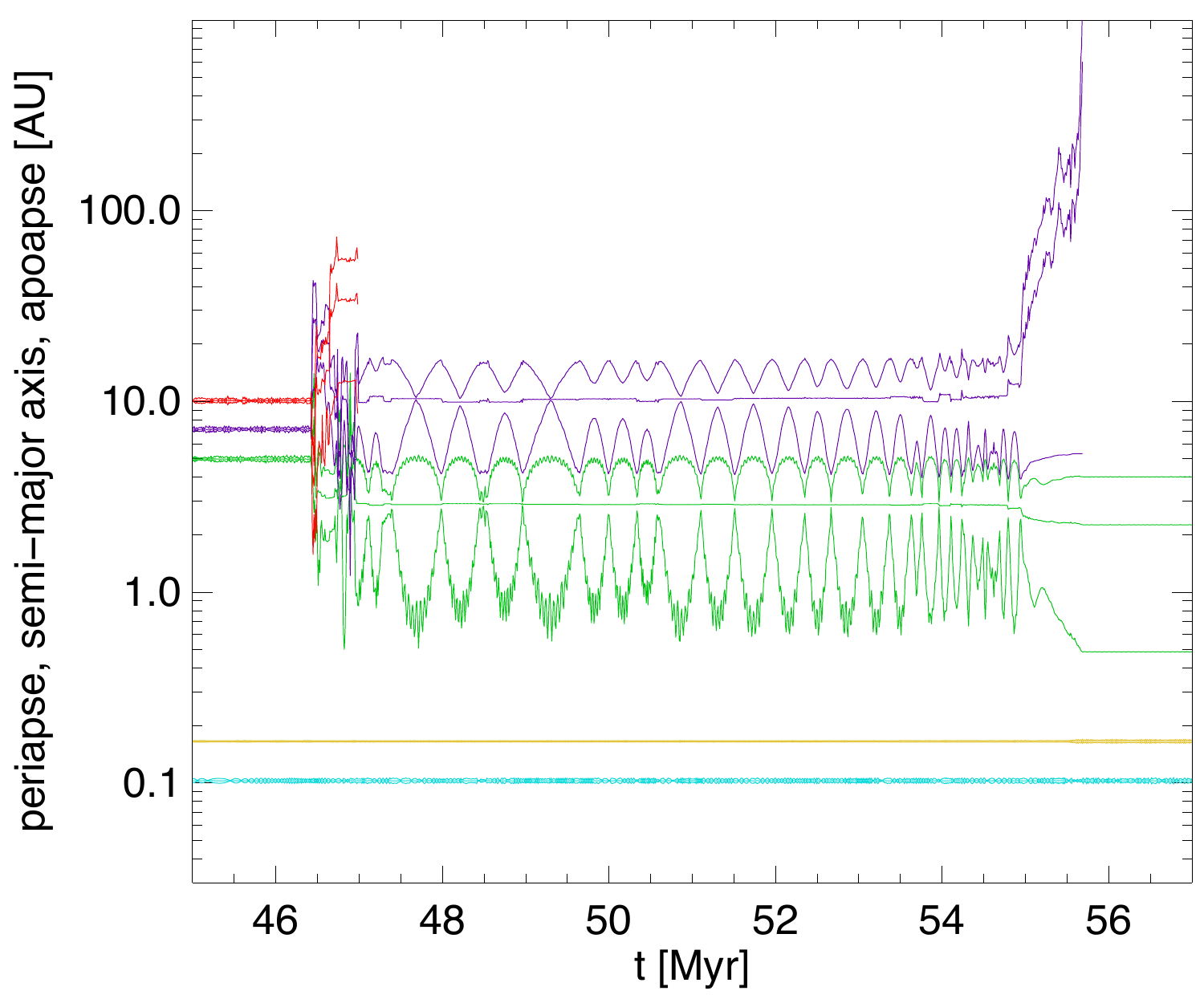} 
\caption{ System S15. Distances as a function of time: the semimajor axis, apastron distance, and periastron distance for each of the planets is shown, for a system that ejects two planets. }
\label{fig:aqq84}  
\end{center}  
\end{figure}

\begin{figure}  
\begin{center}  
\includegraphics[width=0.49\textwidth]{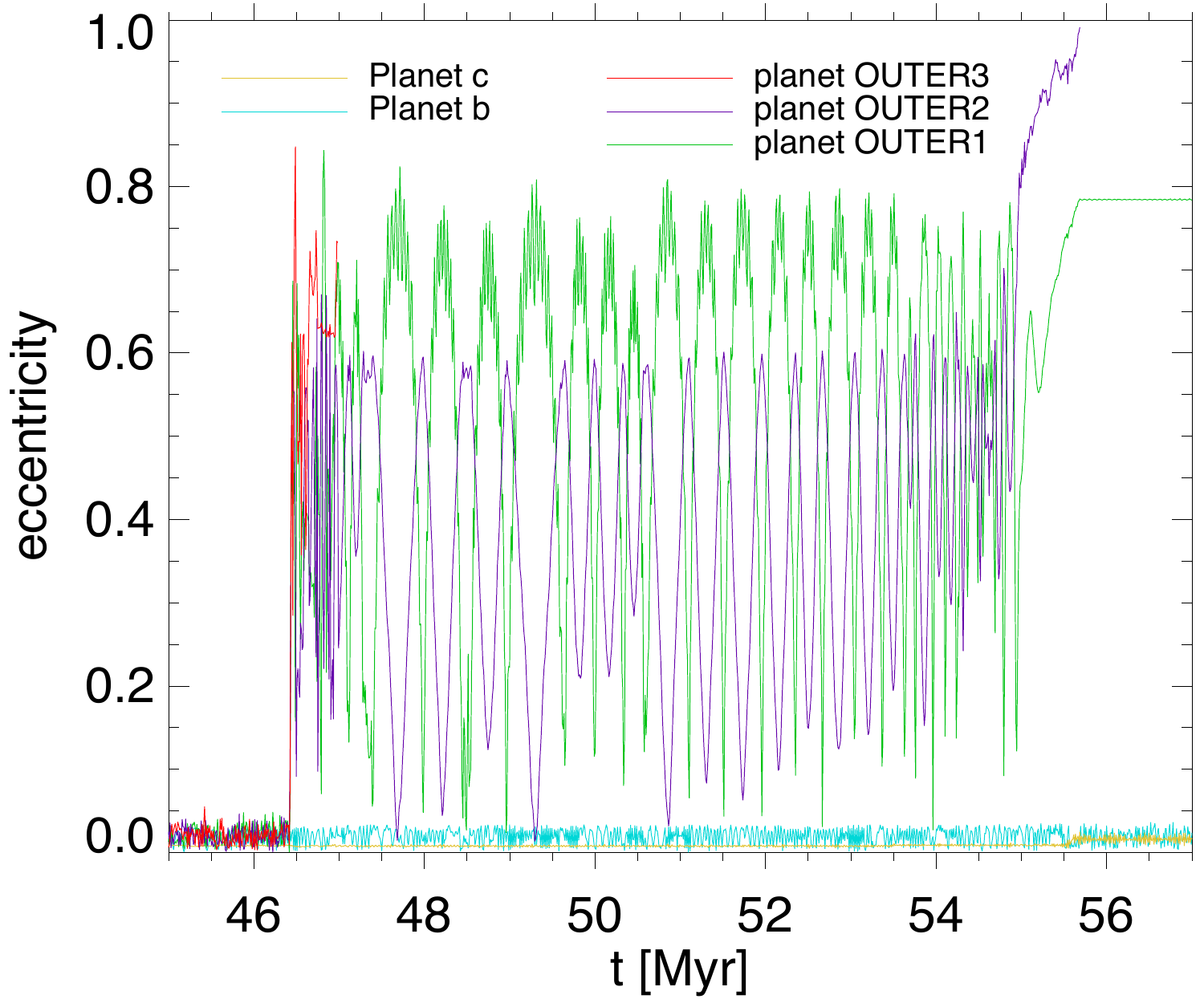} 
\caption{ System S15.  Eccentricity as a function of time. The outer, scattering planets have large fluctuations in eccentricity, whereas the inner ones do not participate in the scattering and remain quite calm. }
\label{fig:ecc84}  
\end{center}  
\end{figure}

\begin{figure}  
\begin{center}  
\includegraphics[width=0.49\textwidth]{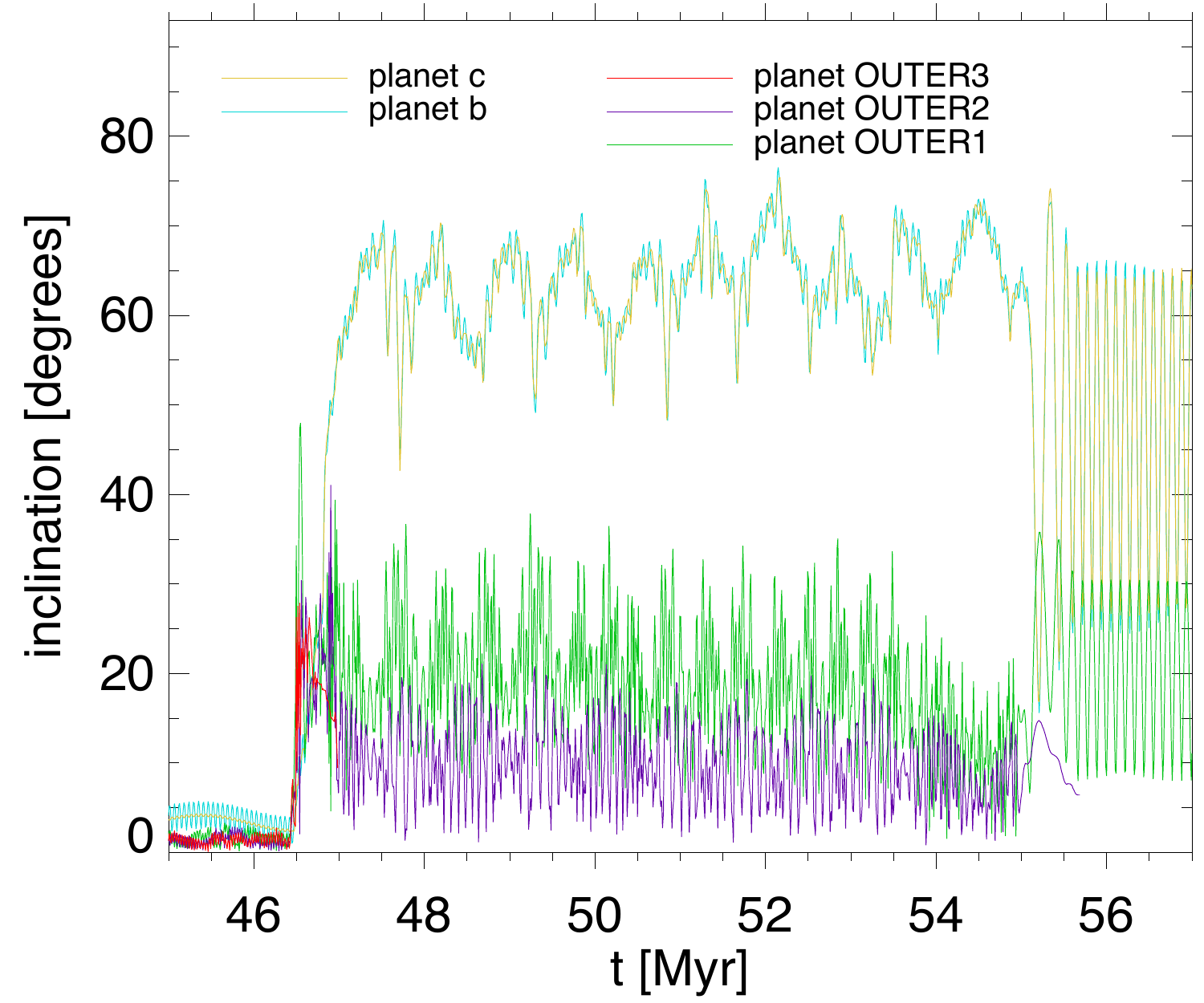} 
\caption{ System S15.  Inclination as a function of time. The inclination is relative to the original plane of the outermost planet. }
\label{fig:inc84}  
\end{center}  
\end{figure}

An example closer to the \cite{2013Huber} hypothesis is shown in figures~\ref{fig:aqq84}, \ref{fig:ecc84}, and \ref{fig:inc84}. A first episode of  scattering over 0.7 Myr leads to the ejection of the outermost planet. The other two outer planets are left eccentric and strongly interacting, although not crossing. After 8 Myr of mostly secular interaction, in a second episode of scattering (again over about 0.7 Myr) the outer planet increases its semimajor axis and eccentricity, then escapes.  The inner planets react to both of those episodes of scattering by quickly changing their modes of inclination variations.  They are left ringing between $25^\circ$ and $65^\circ$, due to the inclination of the final remaining outer planet.  Only in the upper part of that oscillation are the inner planets sufficiently inclined to explain the spin-orbit data.  When the inner planets attain maximum inclination, the now-single outer planet attains its minimum inclination of only $i_d \simeq 8^\circ$.

We note that very spin-orbit misaligned inner planets are not necessarily accompanied by a very spin-orbit misaligned companion planet. In both examples S14 and S15, when the inner planets have inclinations above $45^\circ$, the outer planet in the few-AU region has an inclination $\lesssim15^\circ$.  This property contrasts to the naive prediction of the \citep{2013Huber} hypothesis as discussed in the introduction, that the outer giant would have an inclination $i_d>22.5^\circ$.  There are two effects contributing: (1) a secular resonance effect, in which the forcing frequency of the torque matches the natural precession frequency of the inner planets, amplifying the response (prominently seen in S14, the left panel of figure \ref{fig:inc28}), and (2) the anti-phased oscillations due to precession in the final system (prominently seen in S15, the late-time behavior of figure~\ref{fig:inc84}). 


\section{Statistical results} \label{sec:res}

\subsection{Two Initial Outer-Planets} \label{sec:restwo}
In set 1, we ran a total of $173$ four-planet simulations. The first major event recorded by {\it Mercury} were as follows:
Outer Collisions -- 19, Outer Ejections -- 67, Stable systems -- 67, Inner collisions with star or planet -- 20. 

We analyze the systems that ended up with three planets, after a collision or an ejection occurred. We do not include stable systems, or systems with destroyed inner planets, as they are not acceptable models for the Kepler-56 system. 

Our results indicate that a large-angle (at or above $45^\circ$) spin-orbit misalignment between inner two planets and their host star is highly improbable. We find a total of fourteen systems (out of $173$) for which both inner planets have inclinations between $10^\circ$ and $20^\circ$; the details of these systems are collected in Table~\ref{tab:sys}.  We find an additional seven systems where only one of the two has a larger than $10^\circ$ inclination. Only one system (System S3 in Table~\ref{tab:sys}, detailed in Section~\ref{sec:ex1}) is found to have both inner planets' inclinations around $45^\circ$ that could model Kepler-56, but here the mutual inclination turns out to be large, about $20^\circ$.  However, high mutual inclinations are rare: a total of nine systems --- including S3 --- are found where the mutual inclination exceeds $10^\circ$.  The system with the highest inclinations had its planet initially at $5$ AU -- OUTER1 -- ejected. Generally however, we find that ejections of the outermost planet (OUTER2) are much more common than ejections of the second-outermost planet. 

\begin{table*}
\caption{Properties of our three examples (S1 to S3) with two-outer planets, plus all systems from simulation set 1 with individual inclinations larger than $10^\circ$ and mutual inclination less than $5^\circ$. The final two systems (S14 and S15, italicized) are from systems with three-outer planets initially (simulation set 3).  E = Ejection of either outer planet (all outermost, except S3), C = Collision between outer planets}

\centering 
\begin{tabular}{l | c | c | c | c | c | c | c | c }
\hline\hline 
System        & History & inclination       & inclination     & Mutual  & semi-major & semi-major  &
eccentricity & eccentricity    \\
 &  &  planet b         & planet c       &  inclination & axis b &  axis c &
planet b & planet c     \\
 & & [degrees] & [degrees] & [degrees] & [AU] &[AU] &  &         \\
[0.5ex]
\hline 
Kepler-56 & E or C & $>45$ & $>45$ & $\simeq 5$ & 0.10 & 0.17 & $<0.1$ & $<0.1$ \\
S1 & E & 17.4 & 15.0 & 3.4 & 0.11 & 0.17 & 0.02 & 0.01 \\
S2 & E & 14.7 & 13.7 & 3.4 & 0.10 & 0.17 & 0.02 & 0.02 \\
S3 & E & 55.8 & 46.8 & 20.5 & 0.10 & 0.16 & 0.38 & 0.03  \\
S4 & E  & 12.1 & 10.1 & 2.9 & 0.11 & 0.16 & 0.24 & 0.10  \\
S5 & E & 11.5 & 10.3 & 1.5 & 0.11 & 0.16 & 0.04 & 0.07  \\
S6 & E & 11.4 & 11.6 & 3.5 & 0.10 & 0.16 & 0.03 & 0.01  \\
S7 & E & 17.9 & 19.4 & 1.5 & 0.11 & 0.17 & 0.01 & 0.02\\
S8 & E & 10.8 & 12.4 & 2.2 & 0.10 & 0.16 & 0.04 & 0.02  \\
S9 & E & 16.0 & 15.8 & 2.2 & 0.11 & 0.17 & 0.06 & 0.02  \\
S10 & E & 10.9 & 10.7 & 0.4 & 0.11 & 0.16 & 0.04 & 0.03  \\
S11 & E & 11.4 & 10.6 & 2.1 & 0.10 & 0.16 & 0.03 & 0.02  \\
S12 & E & 10.4 & 11.3 & 3.4 & 0.10 & 0.17 & 0.02 & 0.01  \\
S13 & E & 10.5 & 10.2 & 3.9 & 0.11 & 0.16 & 0.05 & 0.01  \\
\emph{S14} & C & 50-65 & 50-65 & 2.4 & 0.10 & 0.17 & 0.02 & 0.01  \\
\emph{S15} & E,E & 25-65 & 25-65 & 1.8 & 0.10 & 0.17 & 0.02 & 0.01  \\
\hline 
\end{tabular}
\label{tab:sys}
\end{table*}

\begin{figure}  
\begin{center}  
\includegraphics[width=0.49\textwidth]{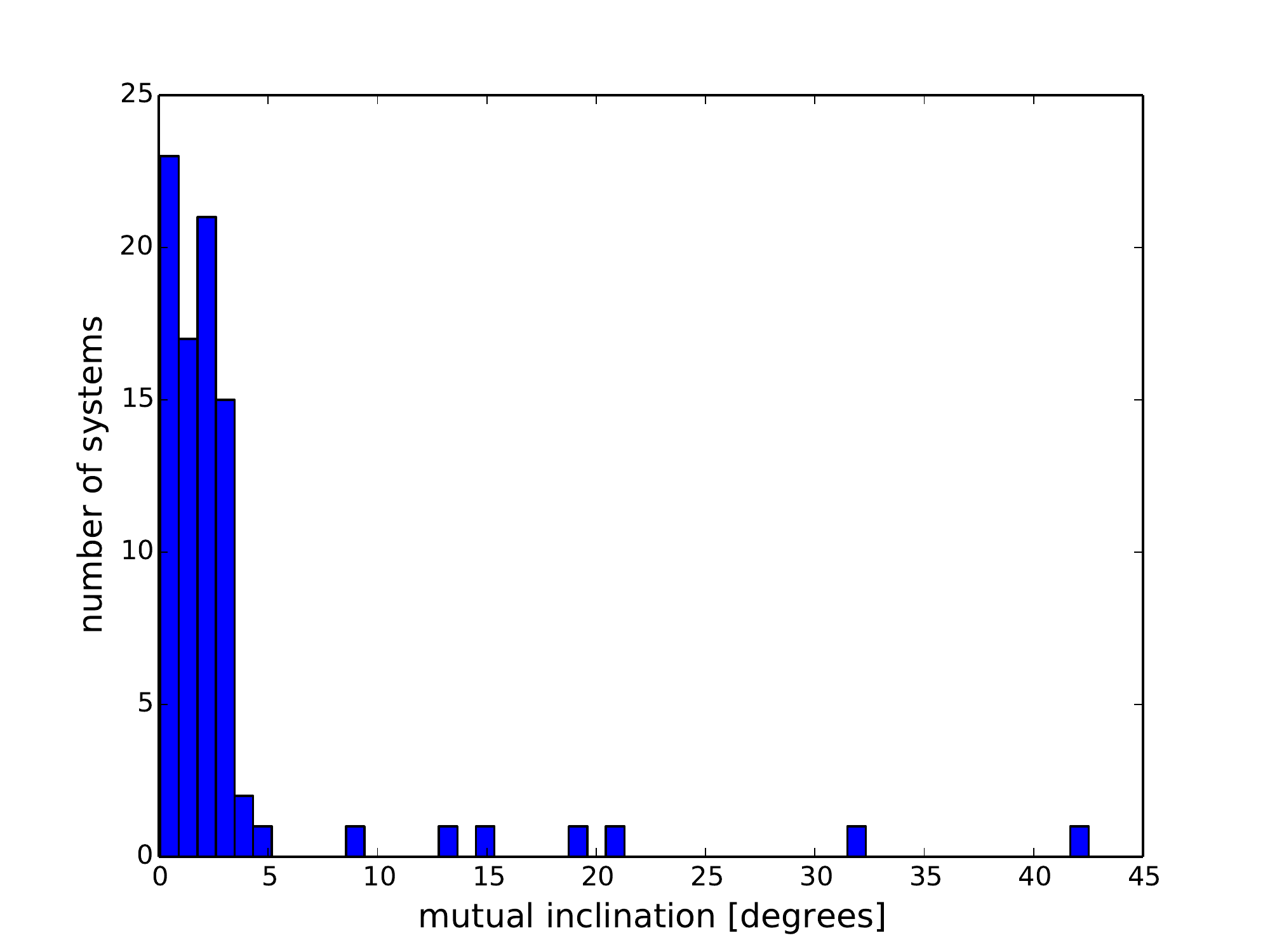} 
\caption{Low final mutual inclinations are common: most simulated systems retain their inner planets' initial coplanarity. However, the one system that ended up with its inner two planets highly inclined belongs to the rare cases of high mutual inclination.}
\label{fig:hist}  
\end{center}  
\end{figure}

\begin{figure} 
\begin{center}  
\includegraphics[width=0.49\textwidth]{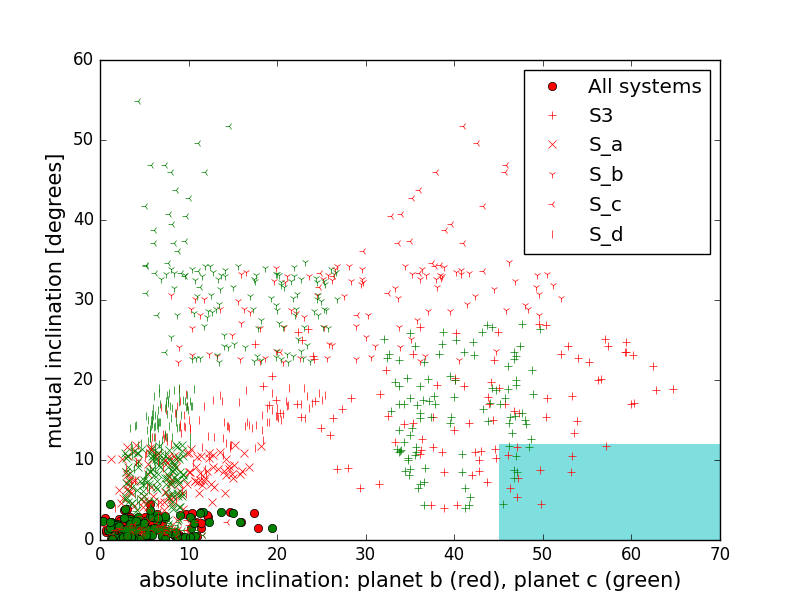} 
\caption{The inner planets' final mutual inclinations are plotted against their final individual inclinations (in red: OUTER1, in green: OUTER2). Systems \texttt{S\char`_a} through \texttt{S\char`_d} (runs left with high mutual inclination) as well as system S3 are plotted at 100 different moments in time: all of those reach equal or higher absolute inclinations and higher mutual inclinations than the rest.  Only S3 achieves both high absolute inclinations while briefly sampling low mutual inclination; this particular system was discussed in section~\ref{sec:s3}. The blue rectangle represents the area containing the current inclinations of the observed Kepler-56 system's inner planets. Most systems' inner planets remain at both low absolute and low mutual inclinations.}
\label{fig:minc}  
\end{center}  
\end{figure}

We see in figure~\ref{fig:hist} that most systems retain a low mutual inclination - a good sign for reproducing the orbital data of Kepler-56. However, what the histogram does not tell us are the absolute inclinations -- this is illustrated in Figure~\ref{fig:minc}, where we show the true mutual inclination $I$ with respect to the absolute inclination of either inner planet.  In fact, the preference for low absolute inclinations of the inner two planets forces us to conclude that four-planet initial conditions does not favor the formation of Kepler-56-type systems.  For systems with large mutual inclination, we integrated 2 Myr forward in time, and show those systems as a cloud of 100 smaller dots sampling this interval in Figure~\ref{fig:minc}.  System S3, despite its usual large mutual inclination, occasionally samples low mutual inclination as well, so it could be an acceptable model for Kepler-56 if observed at a lucky time when it is relatively flat. 

We thus conclude that systems in an initial configuration such as we have chosen, while being able to generate some amount of inclination generically, will only rarely produce systems with both i) high inclinations of both inner two planets, and ii) retaining co-planarity. More common are systems in which only one of the two inner planets was highly inclined, or the inner two planets maintained low absolute inclinations (and as a result, small mutual inclination and low spin-orbit angle).

\subsubsection{Two Planets against Four planets for identical initial parameters}

For simulation set 2, we took $73$ of the systems from set 1, and performed a run without the inner two planets. The idea is to check the two-step concept: the outer planets scatter each other, and later excite the inner planets.  If that is the case, we would expect the branching ratios of the outcomes for the two outer planets to be independent on whether the inner planetary system is in place or not.  One way the branching ratios could differ, for instance, is that the inner planets (having most of the orbital energy) may more readily eject outer planets that come near them on highly eccentric orbits \citep{2015Mustill}. 

We did not find a one-to-one correspondence between individual systems with and without an inner planetary system, presumably due to the chaotic nature of scattering interactions.  Instead, we compare the branching ratios with a statistical test (chi-squared of categorical data). 

We exclude the systems in which the 4-planet version that had one or both of the inner two planets participate in a collision.  That gave 61 systems with the following outcomes: 

\begin{itemize}
\item  Two planets: Outer Collisions -- 16, Ejections -- 23, Stable systems -- 22\\
Four planets: Outer Collisions -- 10, Ejections -- 30, Stable systems -- 21\\
\item Chi-squared: 
\begin{align*}
&\chi^2_2 = 2.33\\
&p = 0.31. 
\end{align*}
\end{itemize}
From these runs, we conclude that the presence of the inner two planets does not statistically affect the outcomes of scattering.  

\subsection{Three Initial Outer-Planets}

\begin{figure} 
\begin{center}  
\includegraphics[width=0.49\textwidth]{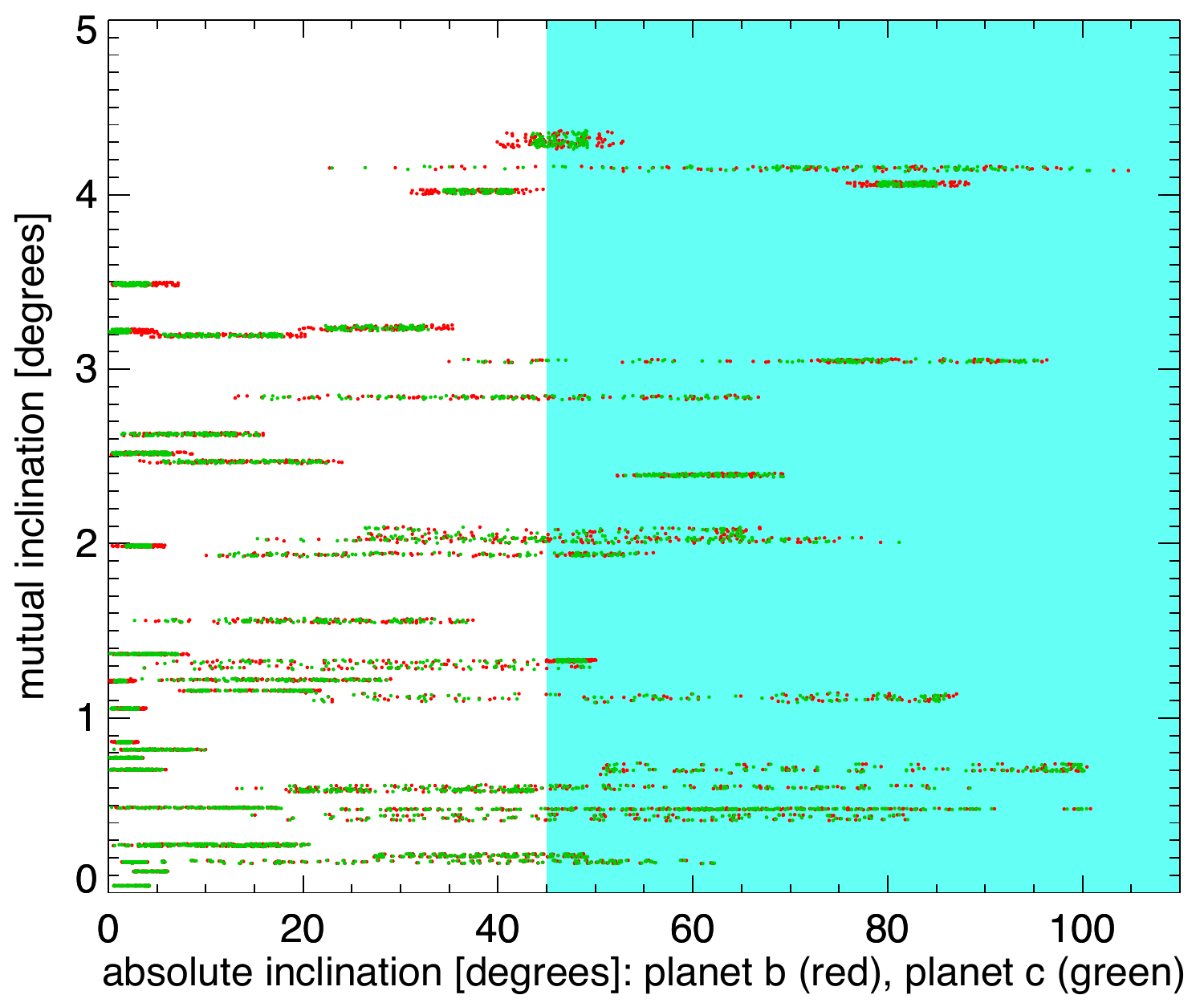} 
\caption{The mutual inclinations are plotted against the individual inclinations of Kepler-56b and Kepler-56c, in the style of figure~\ref{fig:minc}.  Notice the smaller range on the vertical axis with respect to figure~\ref{fig:minc}, as simulation set 3 did not have examples resulting in stable, mutually inclined orbits.}
\label{fig:minc3}  
\end{center}  
\end{figure}

In set 3 of integrations, we found that three outer-planet scattering much more often produces large spin-orbit misalignments.  Of the 100 systems integrated, 28 remained stable $10^8$~years, 27 destroyed one or both inner planets when they went unstable, and 45 went unstable leaving the inner planets intact.  In all these 45 cases, the mutual inclination between the inner two remained low, in contrast to many systems in the two-outer-planet scattering simulations.  For each of these 45 systems, we selected 100 random times after the instability, sampling one or many secular cycles; the values of inclination and mutual inclination are plotted in figure~\ref{fig:minc3}, which is just the three-outer-planet version of figure~\ref{fig:minc}. Of the 45 cases, 19 had inclinations that were above $45^\circ$ or quasi-periodically visited that region. We are presumably seeing the Kepler-56 system at a random phase of its secular evolution, so these points serve as a sample of the hypothesis.  In 28\% of the samples, the planets' individual inclinations are above $45^\circ$.

We looked more closely at the final system architecture of the 19 cases that could explain the Kepler-56 system at some point in their secular cycle.  The most common outcome (13 cases) was ejection of one outer planet, leaving the other two outer planets stable with respect to each other.  In two other cases, two outer planets merged and the merger product was stable with respect to the third outer planet.  So in 15 cases, two outer planets remained.  In the other four systems, one eccentric, outer planet remained: in two cases, two outer planets were ejected; in the other two cases, two of the planets merged and then ejected the other planet. 

In all of these successful cases, a planet of mass 1 or 2 $M_{\rm Jup}$ was left with a semi-major axis between 2.1 and 4.6 AU, and an eccentric orbit. 

In the 15 cases where the inner planets are inclined and a second outer planet remains bound to the system, it does so in a long-period orbit whose signal is inaccessible to Doppler observations.  Quantitatively, in only {\it one} of those runs does this extra planet produce an acceleration greater than 3.2 m/s/yr (see this value's relevance in the conclusions), and it does so only about a quarter of the time (near periastron of an eccentric, 10 AU orbit).  So even though our results suggest planet-planet scattering with at least three initial outer planets is responsible for the misalignment, we only predict an acceleration greater than 1~m/s/yr less than 10\% of the time -- usually in our runs, such an acceleration is even smaller in amplitude, or non-existent. 

\section{Conclusion} \label{sec:concl}
We have run simulations attempting to implement the idea of \cite{2013Huber} for tilting the inner planets via scattering of outer planets, and found it to be unlikely in the case of two-planet scattering but plausible in the case of three-planet scattering. 

We ran $173$ simulations of two outer-planets, of which only one produced high enough spin-orbit misalignment of the inner two planets to match the observations. However, for this outcome, the planet's mutual inclination is about $20^\circ$ for most of its evolution, too high for a reasonable match with Kepler-56 b and c's mutual inclination \citep{2013Huber}. We did not find another system with similarly high inclinations, even though we did find a few systems where one of the inner planets ended up with a high inclination.  We speculate that the system S3 had its second-outermost planet ejected \emph{and} the highest inner-planet inclinations came from the same source -- an epoch of prolonged scattering which allowed access to these dynamically rarer outcomes. 

However, considering our simulations, our hypothesis of the outer planet(s) generating a high spin-orbit misalignment requires particularly violent scattering.  This is apparently possible through 3 equal-mass outer planets, but not with 2.  In runs with three planets in the exterior parts, scattering can tilt the inner system dramatically.  For inner systems that are not disrupted, 28\% showed misalignment from the original plane of the outermost planet by more than $45^\circ$ at a randomly selected time in the future secular evolution of the system.  

Our runs showed that usually two outer planets remain after the scattering, whereas in Kepler-56, only one has been found. 
New data and analysis appears to confirm the existence of a third planet in the Kepler-56 system (Otor, Montet et al., in prep.). It does not exclude the existence of a fourth planet in a large orbit, which could still be hiding. Observationally, a fourth planet has currently an 95\% upper limit on a long term radial velocity acceleration of 3.2 m/s/yr (Otor, Montet et al., in prep), which is why we quoted results with respect to this benchmark above.  The second outer planet in our simulations almost always had a much smaller effect -- it could easily evade that limit. 
 
A useful avenue for future work would be quantifying whether 3 unequal-mass planets can achieve large enough misalignments.  Also, our focus was on one particular system (Kepler-56), but one would rather model a population of systems, whose initial distribution is plausible from planet formation theories, to see what distribution of spin-orbit and mutual inclination outcomes are expected for the inner planets. 

Misalignment of inner planets is likely not rare.  Kepler-56 was the 6th system of multiple transiting planets whose stellar obliquity was measured \citep{2013Albrecht} -- the search turned up an oblique star unexpectedly quickly.  Indeed, another system, KOI-89, has recently been found to feature large angle spin-orbit misalignment of its two inner, coplanar planets \citep{ahlers15}.  In contrast to Kepler-56, there is no known additional object orbiting further out. Nevertheless, it is common enough that if scattering indeed explains this population, then we would suggest multi-planet scattering is more common than two-planet scattering. 

There are observational clues that scattering is probably not the sole mechanism generating misalignments.  \cite{2015Mazeh} have found stellar misalignment to be a strong function of stellar temperature, but not of planetary multiplicity or coplanar architecture.  A third planet does not appear to be needed to produce systems with similar characteristics as Kepler-56, thus weakening scattering as a major mechanism for spin-orbit misalignments of two or more coplanar planets.   A host of other mechanisms may also be in play.  The protoplanetary disk may have been tilted from its inception \citep{1991Tremaine, 2010Bate, 2015Fielding}, or due to magnetic torques in its early stages \citep{2011Lai}.  It could have endured a torque from a previously-bound stellar companion \citep{2012Batygin} or from a flying-by star \citep{2016XiangGruess}.  Even more exotic, the internal convection might have even tilted the stellar surface \citep{2013Rogers} relative to the planetary plane.  Most of these mechanisms would likely leave the non-transiting planet in roughly the same plane as the transiting planets.  Such a configuration will eventually be testable, as orbital precession of the inner planets due to the outer one will become observable in transit data due to the slow but steady duration drifts, the manifestation of planetary precession  \citep{2002MiraldaEscude}.

Our main conclusion is that three outer-planets are necessary for scattering to cause the amount of misalignment inferred for Kepler-56's planets b and c. Two-planet scattering does not seem sufficient, because the excitation is rarely dramatic enough.  Apparently in these cases, scattering in part of the planetary system propagates chaos to all other parts as well.  This conclusion is probably much more general than our attempts to model the Kepler-56 system.  In particular, it has been the upshot of attempts to model the early days of the Solar System \citep{2009Brasser,2012Agnor, 2016Kaib}.  This conclusion may more broadly apply to exoplanets as well.  For instance, since many or most planetary systems of small planets exhibit dynamically packed and rather calm orbits \citep{2014Fabrycky}, and most systems of giant planets have large eccentricity \citep{2010Cumming}, it may suggest that these two types of systems are truly separate, expressing two distinct outcomes of the planet formation process.

\section*{Acknowledgements}

P.G. thanks the National Research Fund Luxembourg for support through grant BFR08-024.  D.F. thanks the Sloan Research Fellowship for support, which funded part of the computer work at the Midway Cluster at the University of Chicago.  

\bibliography{ms.bib} \bibliographystyle{mnras}

\label{lastpage}
\end{document}